\documentclass[letterpaper, 10 pt, conference]{ieeeconf}  

\IEEEoverridecommandlockouts                              

\usepackage{graphics} 
\usepackage{epsfig} 
\usepackage{mathptmx} 
\usepackage{times} 
\usepackage{amsmath} 
\usepackage{amssymb}  
\usepackage{lipsum}
\usepackage{graphicx}
\usepackage{hyperref}
\usepackage{tikz}
\title{\LARGE \bf
Multimodal Deep Learning for Stroke Prediction and Detection using Retinal Imaging and Clinical Data
}

\author{Saeed Shurrab$^{1}$, Aadim Nepal$^{1}$, Terrence J. Lee-St. John$^{2}$, Nicola G. Ghazi$^3$\\
  Bartlomiej Piechowski-Jozwiak$^{3}$, and Farah E. Shamout$^{1}$
\thanks{$^{1}$S. Shurrab, A. Nepal, and F. E. Shamout are with the Division of Engineering, New York University Abu Dhabi, Abu Dhabi, UAE
        {\tt\small saeed.shurrab@nyu.edu, aadim.nepal@nyu.edu, farah.shamout@nyu.edu}}%
\thanks{$^{2}$T. J. L. St. John is with Institute for Healthier Living Abu Dhabi, Abu Dhabi, UAE
        {\tt\small T.leestjohn@ihlad.ae}}%
\thanks{$^{3}$Nicola G. Ghazi is with the Eye Institute at Cleveland Clinic Abu Dhabi, Abu Dhabi, UAE
        {\tt\small GhaziN2@clevelandclinicabudhabi.ae}}%
\thanks{$^{4}$B. Piechowski-Jozwiak is with Canberra Hospital, Canberra, Australia
        {\tt\small bartlomiejpj@gmail.com}}%
}

\newcommand\copyrighttext{%
  \footnotesize \textcopyright Personal use of this material is permitted. Permission from IEEE must be obtained for all other uses, in any current or future media, including reprinting/republishing this material for advertising or promotional purposes, creating new collective works, for resale or redistribution to servers or lists, or reuse of any copyrighted component of this work in other works. 
  This file corresponds to the accepted version of the manuscript published in 47th Annual International Conference of the IEEE Engineering in Medicine and Biology Society.
  DOI: \href{https://doi.org/10.1109/EMBC58623.2025.11253814}{10.1109}}
\newcommand\copyrightnotice{%
\begin{tikzpicture}[remember picture,overlay]
\node[anchor=south,yshift=10pt] at (current page.south) {\fbox{\parbox{\dimexpr\textwidth-\fboxsep-\fboxrule\relax}{\copyrighttext}}};
\end{tikzpicture}%
}

\begin{document}

\maketitle
\copyrightnotice

\thispagestyle{empty}
\pagestyle{empty}

\begin{abstract}
Stroke is a major public health problem, affecting millions worldwide. Deep learning has recently demonstrated promise for enhancing the diagnosis and risk prediction of stroke. However, existing methods rely on costly medical imaging modalities, such as computed tomography. Recent studies suggest that retinal imaging could offer a cost-effective alternative for cerebrovascular health assessment due to the shared clinical pathways between the retina and the brain. Hence, this study explores the impact of leveraging retinal images and clinical data for stroke detection and risk prediction. We propose a multimodal deep neural network that processes Optical Coherence Tomography (OCT) and infrared reflectance retinal scans, combined with clinical data, such as demographics, vital signs, and diagnosis codes. We pretrained our model using a self-supervised learning framework using a real-world dataset consisting of $37$ k scans, and then fine-tuned and evaluated the model using a smaller labeled subset. Our empirical findings establish the predictive ability of the considered modalities in detecting lasting effects in the retina associated with acute stroke and forecasting future risk within a specific time horizon. The experimental results demonstrate the effectiveness of our proposed framework by achieving $5$\% AUROC improvement as compared to the unimodal image-only baseline, and $8$\% improvement compared to an existing state-of-the-art foundation model. In conclusion, our study highlights the potential of retinal imaging in identifying high-risk patients and improving long-term outcomes.
\newline
\indent \textit{Clinical relevance}— This study demonstrates the potential of deep learning models in leveraging retinal images and clinical data for stroke risk prediction and detection. The proposed approach encourages future development of non-invasive, cost-effective technologies for stroke risk assessment. In the long run, this could help mitigate the global stroke burden and improve health outcomes through early intervention. 
\end{abstract}

\section{INTRODUCTION}

Stroke is one of the leading causes of death and long-term disability worldwide, especially among the elderly \cite{feigin2021global}.  The burden of stroke is a pressing global public health problem, with a remarkable increase observed between 1990 and 2019 \cite{feigin2022world}, with low- and middle-income countries accounting for the largest proportion \cite{katan2018global}. This underscores the critical need for robust stroke risk prediction models that enable early prevention among high-risk individuals.

Stroke occurs due to cerebral artery occlusion, also known as ischemic stroke, representing around $80\%$ of cases, or cerebral artery rupture, also known as hemorrhagic stroke \cite{tobin2014neurogenesis}. Common stroke risk factors include history of other disease, such as diabetes, atrial fibrillation, hypertension, and renal failure, or lifestyle factors, such as tobacco use \cite{boehme2017stroke}. Stroke is diagnosed based on the sudden onset of associated symptoms, followed by rapid deterioration of the patient \cite{mosley2007stroke}. This then calls for swift interventions to reduce brain damage and mitigate effects \cite{von2019time}. 

\begin{figure*}[!ht]
    \centering
    \includegraphics[width=\linewidth]{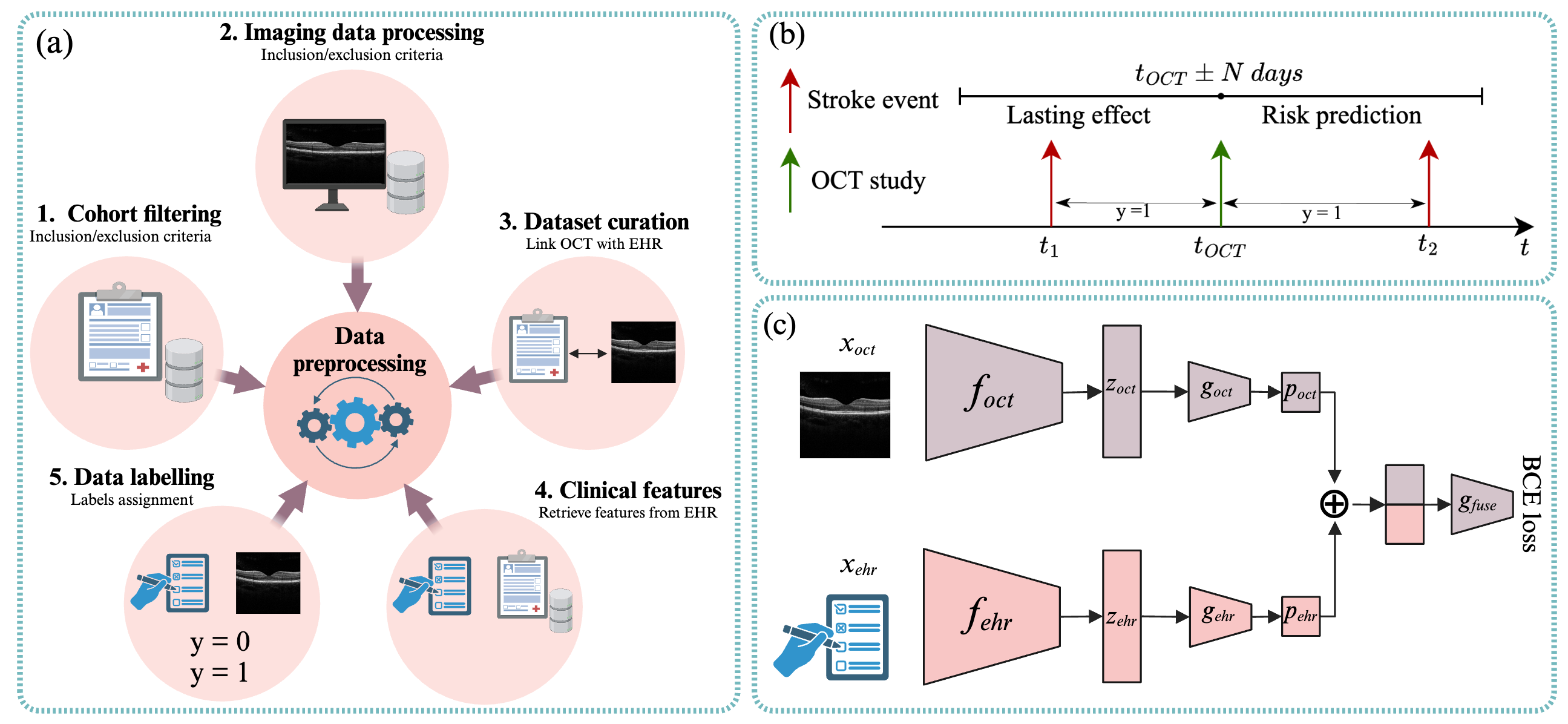}
    \caption{\textbf{Overview of the main components of the RetStroke framework.} (a): Data pre-processing pipeline. (b) Label definition and assignment. (c) Deep neural network architecture.}
    \label{fig:overview}
    \vspace{-4mm}
\end{figure*}

Computed Tomography (CT) and Magnetic Resonance Imaging (MRI) are considered the gold standards for stroke diagnosis following the onset of symptoms \cite{merino2010imaging}. Both techniques are costly, relatively time-consuming, and subject to availability in healthcare centers. This signifies the need for exploring other effective techniques suitable for stroke risk assessment, which are lower in cost and can be quickly acquired in advance of the symptoms onset.

The retina and the brain are known to share anatomical, physiological, and embryological origins \cite{london2013retina}. This renders the retina as a potential candidate for understanding changes occurring in the brain \cite{girach2024retinal}. Furthermore, recent work in neuro-ophthalmology has illustrated that Optical Coherence Tomography (OCT) can encode important information that reflects neural changes in the brain \cite{lamirel2009use}. OCT is a fast, noninvasive, and cost-efficient technique that provides high-resolution in-vivo images of retinal tissue \cite{xie2022use}. Recent studies validated the relevance of OCT imaging in diagnosing various neurological disease \cite{awais2017classification}, such as Alzheimer's. This work falls under the umbrella of an emerging area of research called \textit{oculomics}, which aims to improve our understanding of systemic health and disease through the use of high-resolution retinal imaging and data science techniques \cite{wagner2020insights}. 

Artificial Intelligence (AI), particularly machine learning and deep learning, has significantly advanced healthcare to improve patient diagnosis, prognosis, and treatment. In ophthalmology, the increasing complexity of multimodal datasets and AI techniques offer a strong foundation for exploring the eye-body relationship. For example, several studies focused on the coupled use of retinal images and machine learning for cardiovascular \cite{hu2023systematic} and neuro-degenerative disease \cite{bahr2024deep}. However, existing studies that focus on the analysis of retinal images for stroke using machine learning are relatively limited. Existing work focuses on deriving retinal biomarkers for statistical machine learning models, such as regression and tree-based models, from color fundus photography \cite{messica2024enhancing,yuanyuan2020comparison,zhu2022retinal,rudnicka2022artificial} and OCT angiography \cite{liang2022retinal}. 

A few studies attempted to develop deep learning models for stroke prediction using retinal images. For example, \cite{li2023predicting} trained three different Convolutional Neural Networks (CNN) for one-year stroke risk prediction using fundus photos of different wave lengths. Another work \cite{xiong2024association} developed a CNN using OCT angiography scans to distinguish between retinal images from stroke and non-stroke patients. Lastly, one study \cite{zhou2023foundation} proposed a foundation model (RetFound) for various systemic disease prediction, including ischemic stroke, and trained the model using a large dataset of color fundus photos and OCT separately. The model showed promising performance in predicting three-year stroke risk on the internal validation set but not on the external test set. 

To this end, automated stroke detection using retinal imaging could contribute towards identifying high-risk patients using non-invasive and low-cost modalities, such as OCT. In this study, we propose {RetStroke} (Figure~\ref{fig:overview}), a clinically informed deep learning model that is able to predict stroke using two retinal imaging modalities, specifically OCT and infrared reflectance scans. One novel element of RetStroke is that it makes use of clinically relevant and routinely-collected non-imaging data, such as comorbidities based on diagnosis codes, vital signs, and demographic information during training and inference to enhance predictive performance. Another important aspect of RetStroke is that it is designed for both stroke risk prediction and detection of lasting effects after occurrence of stroke. Finally, despite its simplicity, RetStroke highlights the benefit of multimodal learning and surpasses the performance of an existing state-of-the-art foundation model. 

\section{METHODS}
\vspace{-2mm}
\subsection{Ethics Approval}
This study received all required approvals from the research ethics committees at New York University Abu Dhabi (HRPP-2022-190) and Cleveland Clinic Abu Dhabi (A-2022-059). Informed consent was not required as the study was deemed exempt. 
\vspace{-2mm}

\subsection{Dataset}
We used a private dataset collected at Cleveland Clinic Abu Dhabi (CCAD) between March 2015 and July 2023. CCAD is a multi-specialty large hospital with primary, secondary, and tertiary care facilities in Abu Dhabi, UAE. The dataset includes two modalities, clinical data extracted from the patient EHR and images gathered within an OCT exam. The OCT scans were collected using the Heidelberg Spectralis scanner, which includes spectral-domain OCT scans and infrared reflectance scans, of various eye anatomy such as the retina, peripapillary retina, and the anterior segment. 
The EHR data include patient demographics, diagnoses, vital-sign measurements, and clinical procedures. 

\subsection{Data Preprocessing}
\paragraph{Patient Cohort Filtering} We first specified a set of inclusion and exclusion  criteria in collaboration with domain experts to identify the stroke patient cohort. We restricted our study to adult patients and excluded all patients under the age of $18$ at the time of admission. We included all in-patient encounters that had their admission time and discharge time available, and excluded all other encounter types. Next, we included all patient encounters with a stroke diagnosis code according to the International Classification of Diseases (ICD-10-CM). The target ICD codes were I60 (nontraumatic subarachnoid hemorrhage), I61 (nontraumatic intracerebral hemorrhage), I62 (other and unspecified nontraumatic intracranial hemorrhage), I63 (cerebral infarction), and G459 (transient cerebral ischemic attack, unspecified). 
We then included all patients who underwent CT or CT angiography. Next, we classified the type of stroke as Transient Ischemic Attack (TIA), Ischemic Stroke (IS), and Intracranial Hemorrhage (ICH), according to the diagnosis codes. For patients with TIA, we confirmed the stroke event according to the anti-platelet drug order during the patient stay. For patients with IS, we confirmed the stroke event based on a drug order of recombinant Tissue Plasminogen Activator (r-TPA)  or a dose of anti-platelet drugs. For patients with ICH, we confirmed the stroke event based on a hospital stay longer than $12$ days, according to the standards at the hospital. 

\paragraph{Imaging Data Processing} We also specified a set of inclusion and exclusion criteria for the patients within the OCT dataset. First, we included adult patient and excluded patients below $18$ years. We also restricted our study to patients whose scans captured the macula only. Furthermore, we included all patients with automatic real-time OCT scans and excluded other scanning patterns. We excluded patients whose OCT scans were not associated with infrared reflectance scans. 

\paragraph{Dataset Curation} We used the patient identifiers to curate a dataset that links the OCT studies with the stroke encounters. For each OCT study, we assigned a positive label if the scan was collected within 365 days (one year) from the encounter with a stroke event. We assigned a negative label, if the encounter was not within 365 days from a stroke. We included all negative samples in our dataset to increase the robustness of the model and reflect the real-world distribution where few patients with stroke also have an associated OCT scan. We treated each scan (image) as a single data sample. Finally, we split the dataset into a training and test set with an $80/20$ ratio. We used the patient identifier to split the data to avoid data leakage between splits.

\begin{table}[t]
\caption{Summary of the final study cohort  used for fine-tuning RetStroke.}
\label{cohort-char}
\begin{center}
\resizebox{1.0\linewidth}{!}{
\begin{tabular}{|l|c|c|}
\hline
\textbf{Characteristics} & \textbf{Training Set } & \textbf{Test Set } \\
\hline
\multicolumn{3}{|l|}{\textbf{Demographics}}   \\
\hline 
\ \ \ Number of patients & $5,941$ & $1,486$\\
\hline
\ \ \ Mean age (SD) & $57.8$ ($14.5$)& $58.3$ ($14.4$)\\
\hline
\ \ \ Male patients (\%)& $2,208$ ($37.2$) & $549$ ($36.9$)\\
\hline
 \multicolumn{3}{|l|}{\textbf{Label distribution}} \\
\hline
\ \ \ Patients with stroke (\%) & $139$ ($2.3$) & $44$ ($3.0$) \\
\hline
\ \ \ Studies with positive stroke (\%) & $368$ ($2.4$) &  $100$ ($2.6$) \\
\hline
\multicolumn{3}{|l|}{\textbf{Number of scans}} \\
\hline
\ \ \ Both eyes & $29,833$& $7,754$ \\
\hline
\ \ \ Right eye & $15,467$& $3,988$\\
\hline
\ \ \ Left eye& $14,366$& $3,766$ \\
\hline
\end{tabular}}
\vspace{-8mm}
\end{center}
\end{table}




\paragraph{Clinical Data Processing} We defined and derived a set of static features from the EHR dataset, including demographic and lifestyle factors, vital-sign measurements, and history of disease based on ICD codes (total of 34 features). The first category included patient age, sex, and smoking status. The vital signs included body mass index, systolic blood pressure, diastolic blood pressure, temperature,	pulse rate, and	respiratory rate. We used the vital signs collected from the same OCT encounter, if present, otherwise we used these measurements collected in the most recent encounter preceding the OCT visit. If vital-sign measurements were not available, we replaced them with standard `normal' values based on clinical knowledge. History of disease was represented using the ICD code groups. We retrieved all diagnoses codes assigned to the patient prior to the OCT visit. All categorical variables were one-hot encoded, whereas the numerical variables were min-max normalized.  

\paragraph{Data Labeling for Training and Evaluation} We formulated the model prediction task as a binary classification task using the labels assigned to the OCT scans. It should be noted that our labeling procedure considers both situations in which a stroke event occurs before or after OCT scan acquisition. We trained and evaluated our model using this overall label considering the limited size of the dataset in such a retrospective study. 

Assume $t_{OCT}$ is the time of OCT scan acquisition and $t_{stroke}$ is the time of stroke occurrence, we also evaluated the model for two more scenarios: risk prediction, where the stroke occurs after the OCT ($t_{stroke} > t_{OCT}$), and detection of lasting effects, where the stroke occurs before the OCT ($t_{stroke} < t_{OCT}$). For a more granular analysis, we assessed model performance across different time horizons ($N$ days), including 90, 180, 270, and 365 days.



\subsection{Prediction Model}

\paragraph{Problem Formulation}
Consider that $\mathcal{D} = \{(x_{oct}^{i}, x_{ehr}^{i}, y^{i}) \mid i = 1, 2, \dots, n\}$ is a multimodal labeled dataset, where $n$ is the number of samples. Let $x^{i}_{oct}\in \mathbb{R}^{h\times w}$ represent an OCT scan, where $h$ and $w$ are the height and width of the image. Assume the image is associated with $x^{i}_{ehr} \in \mathbb{R}^{s}$, which is defined as a vector of static features derived from the patient's EHR data, where $s$ is the number of features. Consider that each data tuple ($x^{i}_{oct}$, $x^{i}_{ehr}$) is associated with the ground-truth label $y^{i} \in \{0,1\}$, which represents the stroke label. Our goal is to train a multimodal neural network that can predict $y^i$ given the input data.

\paragraph{Architecture}
RetStroke is a multimodal neural network designed to process imaging data and clinical information. Our proposed model consists of four modules: (i) a visual encoder parametrized as a CNN and denoted by $f_{oct}$, (ii) an EHR encoder parametrized as a multi-layer perceptron network and denoted by $f_{ehr}$, (iii) a non-parametric fusion module denoted as $\oplus$, and (iv) a prediction head parametrized as a fully connected layer, denoted by $g_{fuse}$. We provide further details to clarify the role of each module. 

First, the OCT image, $x_{oct}$, is encoded by the visual encoder $f_{oct}$ to obtain the representation $z_{oct}$. We also encode the static EHR data, $x_{ehr}$, via the EHR encoder $f_{ehr}$ to obtain the representation $z_{ehr}$. RetStroke utilizes the \textit{late fusion} strategy to combine information from both modalities. Hence, both the visual encoder and EHR encoder have their own prediction heads denoted as $g_{oct}$ and $g_{ehr}$, respectively, which process $z_{oct}$ and $z_{ehr}$ to obtain  modality-specific predictions $p_{oct}$ and $p_{ehr}$. Both prediction are then concatenated via the fusion module $\oplus$ and passed to RetStroke's main prediction head, $g_{fuse}$, to obtain the final prediction $\hat{y}$.  We used the Binary Cross-Entropy (BCE) loss function to train the model:
\vspace{-1mm}
\begin{equation}
    \label{bce}
    \mathcal{L} = - \frac{1}{n} \sum_{i=1}^n \left[ y_i \log(\hat{y}_i) + (1 - y_i) \log(1 - \hat{y}_i) \right].
\end{equation}

\paragraph{Training Strategy} As we had a relatively small labeled dataset to train RetStroke, we adopted a two-stage training strategy: self-supervised pre-training of the visual encoder with unlabeled data followed by downstream fine-tuning with the labeled dataset. For self-supervised pre-training, we adopted SimCLR \cite{chen2020simple}, which is an unsupervised pre-training framework that uses the contrastive loss to learn high quality representations. The contrastive loss allows the model to learn representations by maximizing the similarity between positive pairs and minimizing it with negative pairs:
\vspace{-1mm}
\begin{equation}
    \label{contrastive}
    \mathcal{L}_{i,j} = -\log\frac{\exp\left(\text{sim}\left(\mathbf{z}_{i}, \mathbf{z}_{j}\right)/\tau\right)}{\sum^{2K}_{k=1}1_{[k\neq{i}]}\exp\left(\text{sim}\left(\mathbf{z}_{i}, \mathbf{z}_{k}\right)/\tau\right)},
\end{equation}

where, K is the number of examples in a mini-batch, $\mathbf{z}$ is the latent representation computed by the visual encoder, and $\tau$ is the temperature parameter that controls the sharpness of the similarity scores computed between pairs. Instead of fixing $\tau$, we use a learnable temperature parameter whose value is adjusted automatically during training. 

For contrastive model pre-training, we developed two OCT datasets and one infrared reflectance dataset. Since the OCT scan is of volumetric nature, where a single eye scan contains multiple slices ranging in $(25-49)$ scans, we treated the slices as independent samples, which resulted in a dataset containing $1.1$ million image. The second pre-training OCT dataset consisted of the mid-slice only of each OCT volume, while the infrared reflectance dataset consisted of a single image from each study in the dataset. We split the pre-training datasets based on ratio of $90/10$ training/validation splits. After pretraining, we load the pre-trained weights to initialize the visual encoder encoder, $f_{oct}$, of RetStroke and we fine-tuned the entire architecture using the stroke labels.



\vspace{-1mm}
\subsection{Experimental Setup}
\paragraph{Model Pre-training}
 To setup SimCLR for pretraining, we followed the same augmentations used in the original paper, including random resized crop, color jittering, Gaussian blur, and random vertical and horizontal flip (referred to as \textit{harsh} augmentations). We also computed dataset specific statistics and use them for data normalization. We used \textit{AdamW} \cite{loshchilov2017decoupled} 
 as an optimizer and a batch size of $256$.  We also used \textit{cosine annealing} for the learning rate scheduler. For $\tau$, we set the initial value as $0.5$ and fix it as $0.1$ if its value goes below this threshold during training. We pre-train each model for $200$ epochs with early stopping, using a patience of $10$ epochs and a minimum change of $1e-6$ in the validation loss. We also optimize the learning rate and weight decay with values in the intervals $[1e-6, 1e-5]$ and $[0, 1e-1]$, respectively, for the pre-training experiments using random search with 10 runs for each experiment.

\paragraph{Model Fine-tuning}
We used ResNet-18 
as the backbone for the visual encoder of RetStroke, $f_{oct}$. We used two fully connected layers with batch normalization and ReLU activation for the EHR encoder, $f_{ehr}$. For the prediction head $g_{fuse}$, we used a linear layer that processes the fused predictions of both encoders to compute the final prediction. We used the \textit{Adam} optimizer 
and \textit{cosine annealing} for the learning rate scheduler. For image augmentation, we applied random horizontal flipping, rotation, shearing and translation with probability of $0.5$ (referred to as \textit{simple} augmentations).

To optimize the hyperparameters of RetStroke, we performed Bayesian hyperparameter search with K-fold cross-validation. We used the Area Under the Receiver Operating Characteristic curve (AUROC) to select the best models using the validation set. We set the number of folds to five in all experiments. The optimized hyperparameters include the learning rate with values in $[1e-6,5e-5]$, weight decay with values in $[0,1e-3]$, batch size with values in $[64,128,256]$, and augmentations with values in [\textit{simple}, \textit{harsh}]. We also set the number of epochs during hyperparameter tuning to $100$ with a patience of $10$ epochs. We performed $100$ optimization trials for each experiment. After hyperparameter tuning, we selected the model with the best cross-validation AUROC and evaluated on the test set. We used Nvidia A100 GPUs for all experiments.


\paragraph{Baselines} To assess the robustness of RetStroke, we compared our results with two baselines: (i) unimodal image-only training and (ii) the RetFound foundation model. RetFound is a foundation model trained with a large dataset that is capable of performing multiple prediction tasks. We froze its weights and used it as a feature extractor to maintain fairness of comparisons. We also conducted an experiment with RetFound where we incorporated the clinical features utilized by RetStroke to investigate its performance in multimodal settings. Lastly, we utilize the same settings used in our experiments to train RetFound. For model evaluation, we report AUROC, Area Under the Precision Recall Curve (AUPRC), and sensitivity at fixed specificity.

\section{RESULTS}
After applying the inclusion and exclusion criteria to identify encounters with a stroke, the patient cohort consisted of $5,523$ patients diagnosed with stroke, who were associated with $6,152$ hospital encounters. The raw OCT dataset was collected from $9,056$ patients consisting of $23,708$ unique studies. After applying the inclusion and exclusion criteria to the OCT dataset, the OCT dataset consisted of $18,996$ unique OCT studies collected from $7,427$ patients. Linking both datasets resulted in $468$ positive OCT studies, collected from $183$ patients diagnosed with stroke. The final dataset consisted of $37,587$ scans for both eyes, with $19,455$ scans of the right eye and $18,132$ of the left eye. Table \ref{cohort-char} summarizes the main characteristics of the final dataset used for fine-tuning RetStroke.



\begin{figure}[!t]
    \centering
    \includegraphics[width=1.0\linewidth]{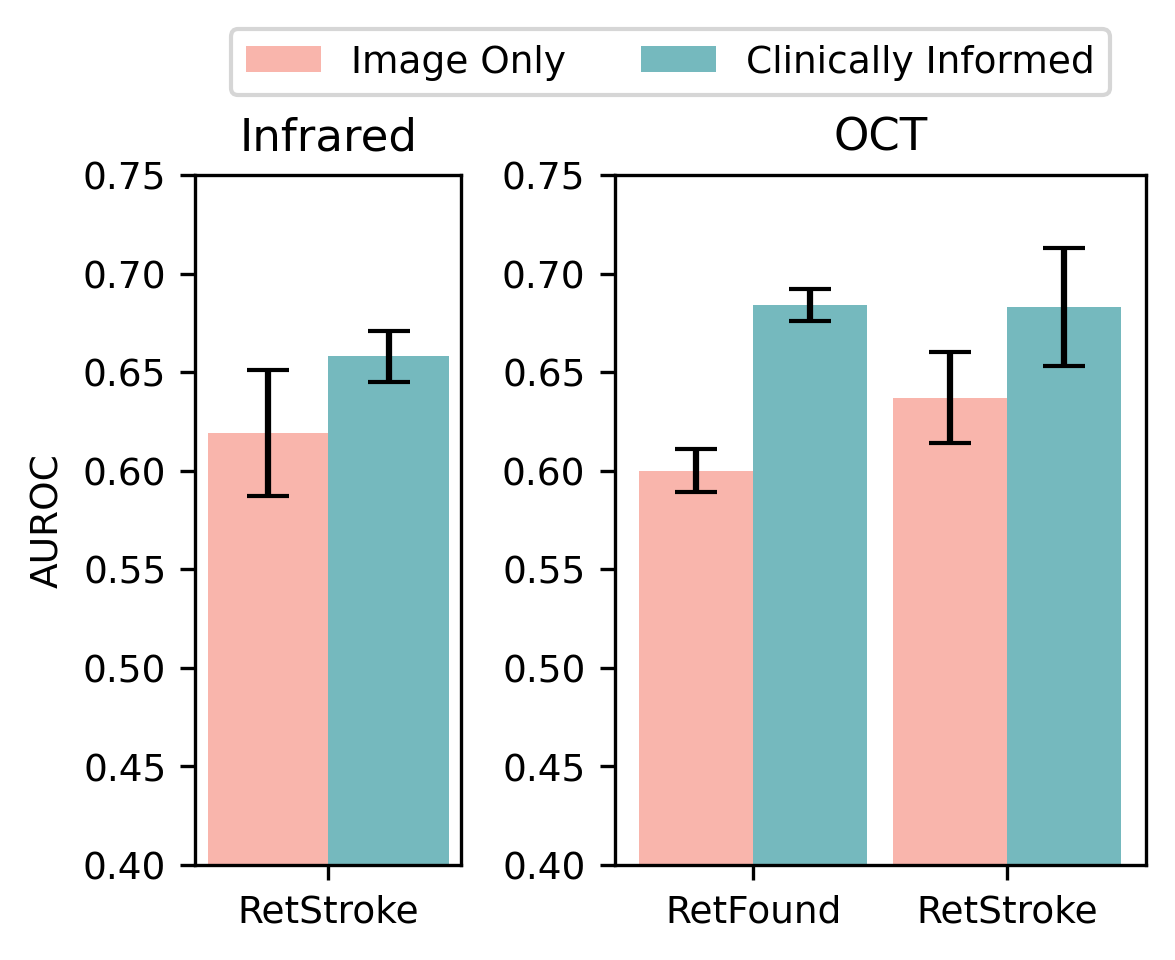}
    \caption{{Overall performance for unimodal and multimodal models.}}
    \label{fig:overall}
\end{figure}


\begin{table}[!ht]
\centering
\caption{Overall AUROC performance ($\pm$ Standard Deviation) compared across unimodal and multimodal baselines.}
\label{tab:overall}
\begin{tabular}{|ccc|}
\hline
\multicolumn{1}{|c|}{\textbf{Models}} & \multicolumn{1}{c|}{\textbf{Unimodal}} & \textbf{Multimodal} \\ \hline
\multicolumn{3}{|c|}{\textbf{Infrared}}                             \\ \hline
\multicolumn{1}{|c|}{\textbf{RetStroke}} & \multicolumn{1}{c|}{0.619 (0.032)} &  \textbf{0.658 (0.013)}\\ \hline
\multicolumn{3}{|c|}{\textbf{OCT}}                                  \\ \hline
\multicolumn{1}{|c|}{\textbf{RetFound}}  & \multicolumn{1}{c|}{0.600 (0.011)} & \textbf{0.684 (0.01)} \\ \hline
\multicolumn{1}{|c|}{\textbf{RetStroke}} & \multicolumn{1}{c|}{\textbf{0.637 (0.023)}} & 0.683 (0.03) \\ \hline
\end{tabular}
\end{table}

Table \ref{fig:overall} summarizes the performance of RetStroke in the unimodal and multimodal settings for both infrared and OCT in comparison with RetFound for the latter, whereas \ref{fig:overall} depicts the results visually. Using infrared scans, RetStroke achieves AUROC of $0.658$ compared to its unimodal variant which achieves an AUROC of $0.619$. For OCT scans, the unimodal variant of RetStroke achieves a better performance than RetFound (0.637 vs 0.600 AUROC). Furthermore, RetStroke and RetFound benefit from the incorporation of clinical information, i.e. multimodal setting. 

\begin{figure}[t]
  \centering
  \includegraphics[width=0.93\linewidth]{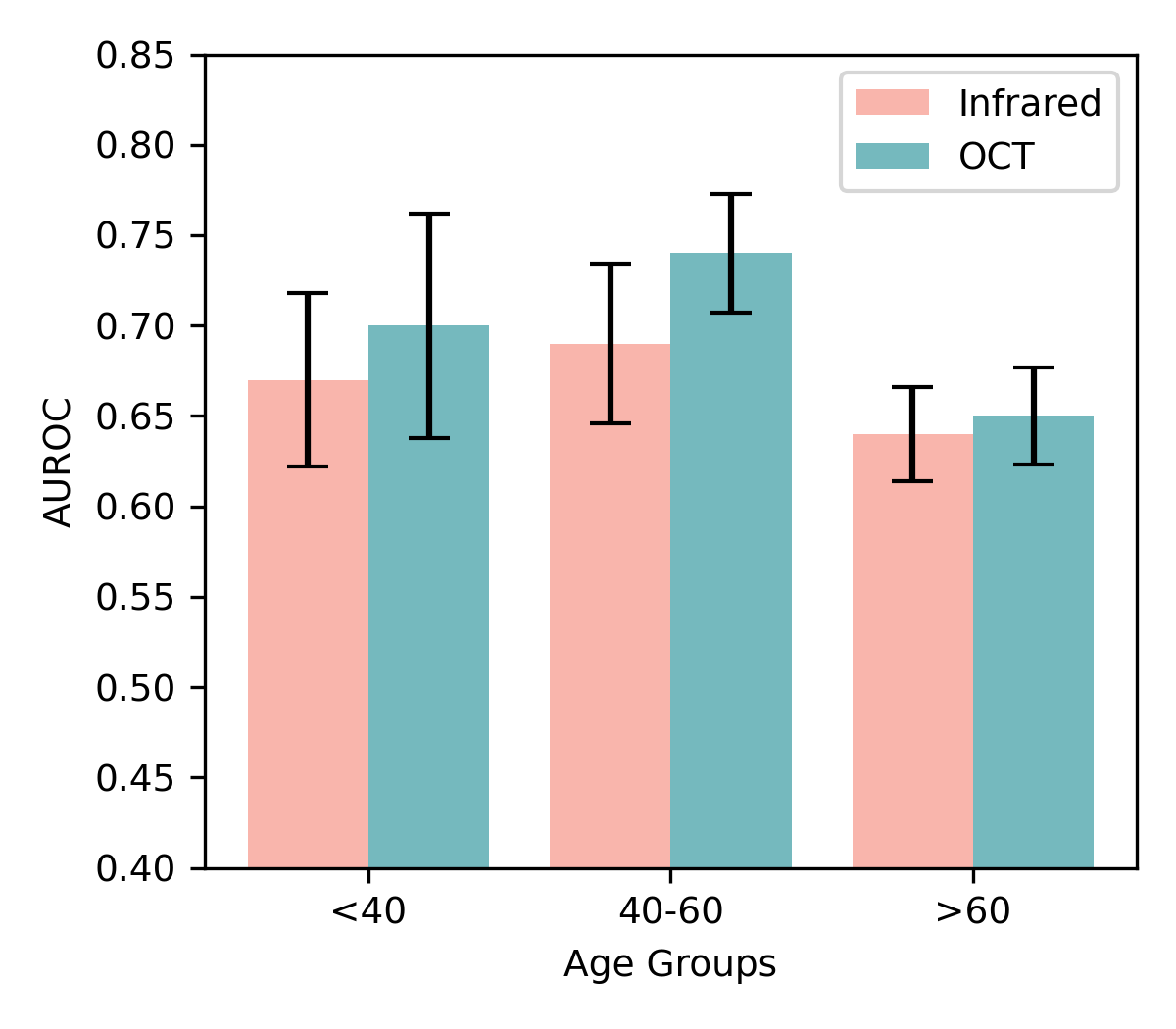}
  \caption{{Performance of RetStroke based on age groups.}}
  \label{fig:age-groups}
\end{figure}

To better understand the performance of our proposed model, RetStroke, we conduct a subgroup analysis considering patient age, comorbidities, and stroke subtype. As shown in Figure \ref{fig:age-groups}, RetStroke achieves the best performance in the $(40-60)$ years subgroup, with an AUROC score of $0.740$ and $0.690$ for OCT and infrared, respectively. For the younger age group $(<40)$, the performance of RetStroke slightly degrades, achieving AUROC score of $0.700$ and $0.670$ for OCT and infrared, respectively. Finally, RetStroke shows a relatively low performance with respect to the elderly age group $(>60)$ compared to the other age groups, around $0.640$. We note that the mean age of the patient cohort is 57.8.

\begin{figure}[t]
  \centering
  \includegraphics[width=0.91\linewidth]{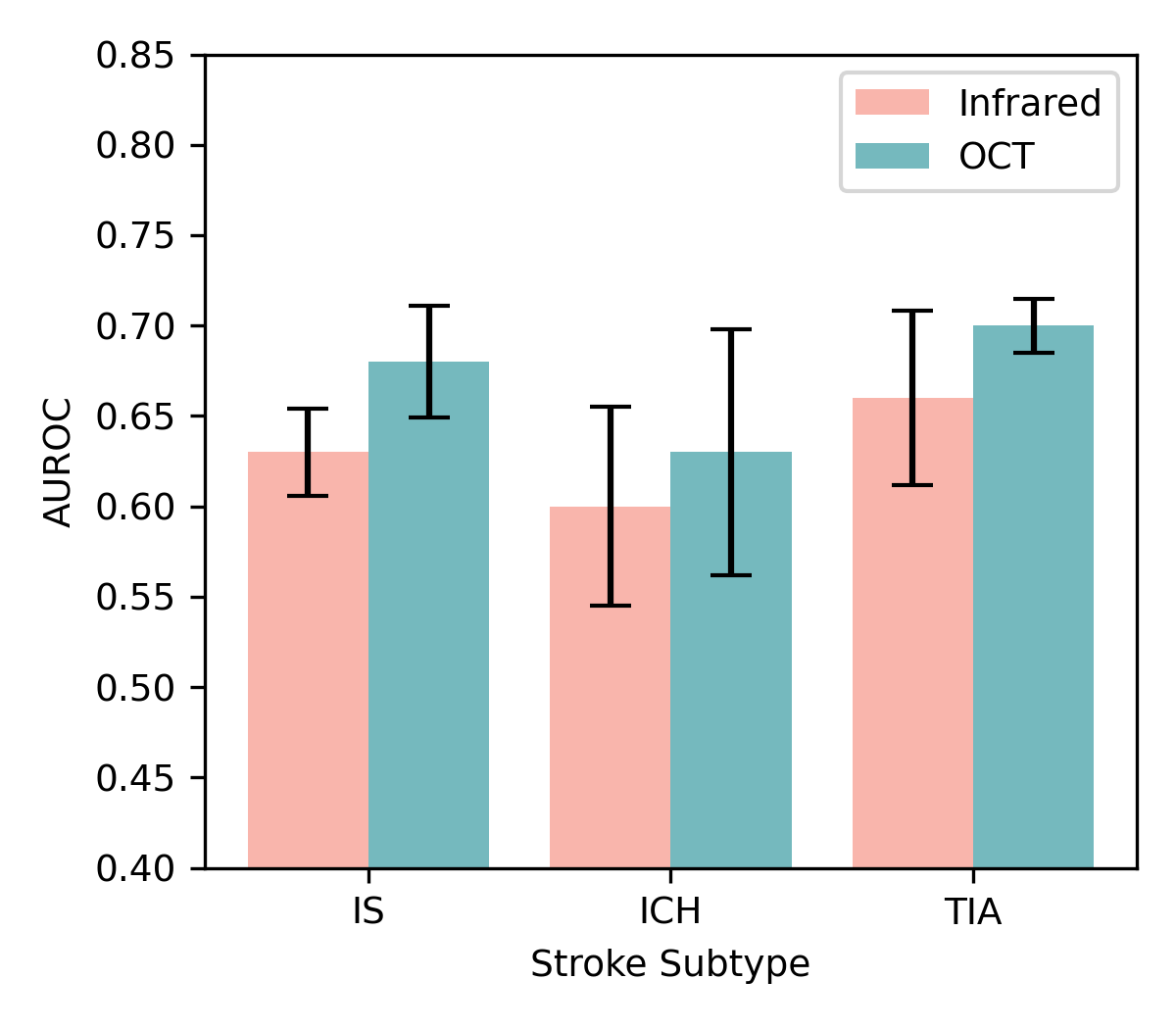}
  \caption{{Performance of RetStroke based on Stroke subtype} (IS: Ischemic Stroke, ICH: Intracranial Hemorrhage, and TIA: Transient Ischemic Attack).}
  \label{fig:subtype}
\end{figure}

Figure \ref{fig:subtype} depicts RetStroke performance with respect to stroke subtypes. We observe that RetStroke performs better with respect to ischemic subtypes, including IS and TIA, with an AUROC score around $0.710$ for both subtypes using OCT and $0.650$ using infrared. The model achieves lower AUROC scores for ICH ($0.61-0.63$). This could be justified by the high prevalence of ischemic stroke ($80\%$) compared to the hemorrhagic stroke ($20\%$).

\begin{figure}[t]
  \centering
  \includegraphics[width=\linewidth]{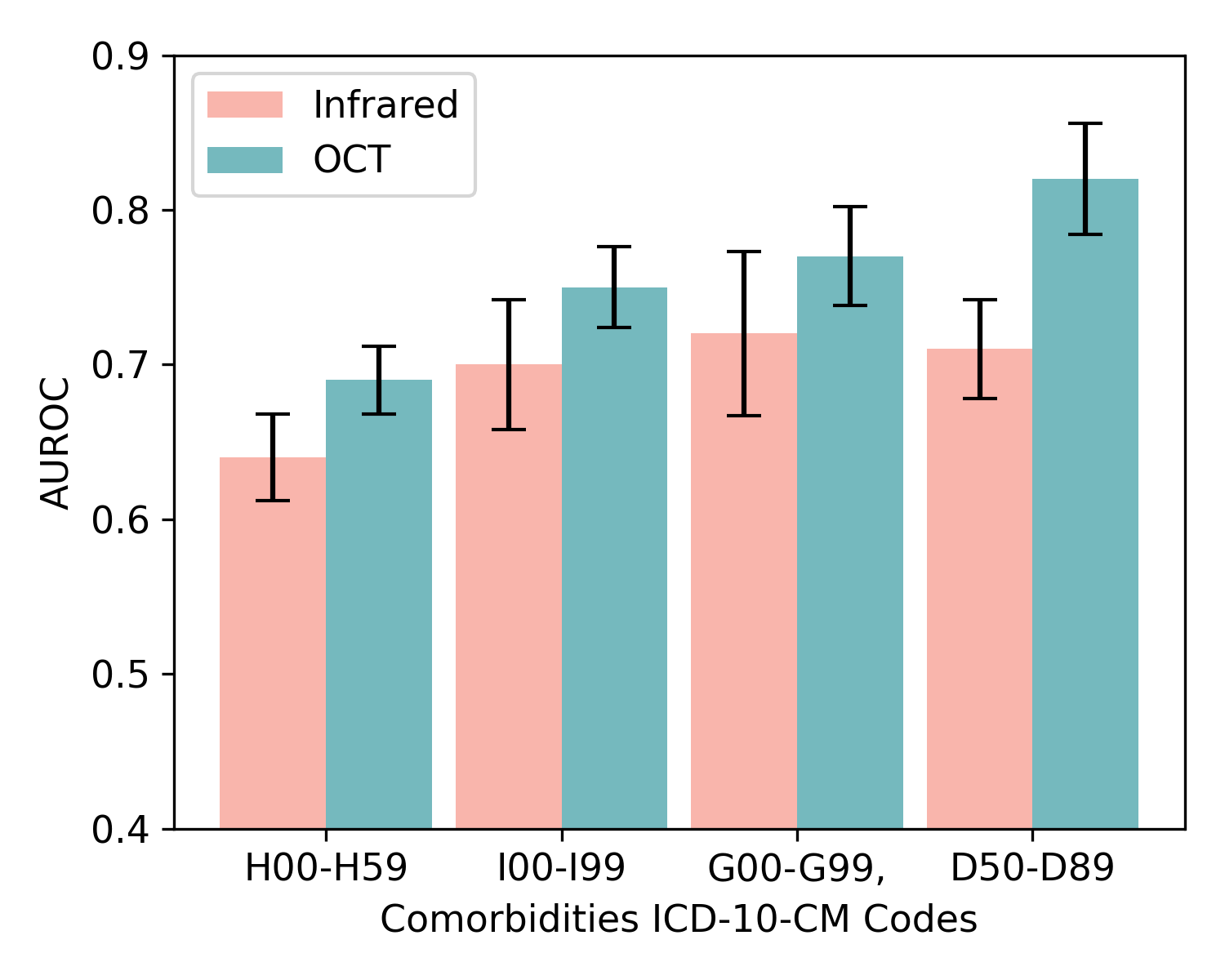}
  \caption{{Performance of RetStroke based on comorbidities, for H00-H59 (diseases of the eye and adnexa), I00-I99 (diseases of the circulatory system), G00-G99 (diseases of the nervous system), and D50-D89 (diseases of the blood and blood-forming organs and certain disorders).}}
  \label{fig:comorb}
\end{figure}

Figure \ref{fig:comorb} illustrates RetStroke's performance concerning patients' historical comorbidities prior to scan acquisition. The best performance is observed among patients with blood-related diseases (D50-D89) using OCT scans, achieving an AUROC score of 0.819, while performance is significantly lower with infrared scans (0.711) for the same comorbidity. For nervous system (G00-G99) and circulatory system (I00-I99) disease, RetStroke achieves similar performance, with AUROC scores around 0.760 using OCT scans history of stroke-related diseases. Conversely, RetStroke exhibits the lowest performance among patients with eye disease (H00-H59), with AUROC scores of 0.686 using OCT scans and 0.640 using infrared scans. This finding aligns with previous research indicating that the presence of ophthalmic diseases degrades the model's predictive performance \cite{girach2024retinal,diaz2022predicting}.

We also investigated the performance of RetStroke with respect to stroke prediction and detection of lasting events, as shown in Table \ref{tab:assoc/risk}. With infrared scans, RetStroke achieves a higher performance in stroke risk prediction especially in the first six months, with AUROC of $0.774$ and $0.766$ for the time periods $<90$ and $<180$, respectively. As the time period increases, performance remarkably decreases, with AUROC reaching below $0.710$ for $<270$ and $<365$ days. On the other hand, the ability of RetStroke to detect lasting effects of stroke is relatively lower with infrared compared to risk prediction, with an AUROC ranging between $0.620$ and $0.640$.

With the OCT scans, RetStroke achieves an AUROC of $0.723$  and $0.736$ for risk prediction within $<90$ and $<180$ days, respectively. The performance similarly decreases within the longer time horizons with AUROC scores dropping below $0.71$. For the detection of lasting effects, RetStroke has a similar performance across the first three time horizons with AUROC $0.705-0.719$.. One interesting observation is that RetStroke generally performs better in the risk prediction task using the infrared modality. While for lasting effect, RetStroke performs significantly better with OCT than infrared across all time horizons.

Finally, in Table~\ref{tab:sens/spec}, we present the sensitivity of RetStroke for risk prediction across infrared and OCT for 0.5 specificity. The model achieves a relatively better performance for OCT than infrared.

\begin{table}[!t]
\caption{RetStroke AUROC performance  ($\pm$ Standard Deviation) for stroke risk prediction and detection of lasting effects across different time horizons (in days).}
\label{tab:assoc/risk}
    \centering
    \resizebox{1.0\linewidth}{!}{
\begin{tabular}{|c|cccc|}
\hline
\textbf{Modality} &
  \multicolumn{1}{c|}{\textbf{\textless{}90}} &
  \multicolumn{1}{c|}{\textbf{\textless{}180}} &
  \multicolumn{1}{c|}{\textbf{\textless{}270}} &
  \textbf{\textless{}365} \\ \hline
\multicolumn{5}{|c|}{\textbf{Detection of lasting effects}}                                                                                    \\ \hline
\textbf{Infrared}  & \multicolumn{1}{c|}{0.627 (0.032)} & \multicolumn{1}{c|}{0.621 (0.026)} & \multicolumn{1}{c|}{0.624 (0.033)} & 0.639 (0.034) \\ \hline
\textbf{OCT}        & \multicolumn{1}{c|}{0.705 (0.055)} & \multicolumn{1}{c|}{0.719 (0.058)} & \multicolumn{1}{c|}{0.719 (0.057)} & 0.735 (0.054) \\ \hline
\multicolumn{5}{|c|}{\textbf{Risk prediction}}     \\ \hline
\textbf{Infrared} & \multicolumn{1}{c|}{0.774 (0.034)} & \multicolumn{1}{c|}{0.766 (0.036)} & \multicolumn{1}{c|}{0.665 (0.045)} & 0.708 (0.046) \\ \hline
\textbf{OCT}       & \multicolumn{1}{c|}{0.723 (0.069)} & \multicolumn{1}{c|}{0.736 (0.045)} & \multicolumn{1}{c|}{0.684 (0.021)} & 0.707 (0.018) \\ \hline
\end{tabular}}
\end{table}

\begin{table}[!t]
\caption{RetStroke sensitivity@0.5 specificity ($\pm$ Standard Deviation) for stroke risk prediction across different time horizons (in days).}
\label{tab:sens/spec}
    \centering
    \resizebox{1.0\linewidth}{!}{
\begin{tabular}{|c|cccc|}
\hline
\textbf{Modality} &
  \multicolumn{1}{c|}{\textbf{\textless{}90}} &
  \multicolumn{1}{c|}{\textbf{\textless{}180}} &
  \multicolumn{1}{c|}{\textbf{\textless{}270}} &
  \textbf{\textless{}365} \\ \hline
\textbf{Infrared} & \multicolumn{1}{c|}{0.655 (0.356)} & \multicolumn{1}{c|}{0.627 (0.383)} & \multicolumn{1}{c|}{0.576 (0.406)} & 0.600 (0.389) \\  \hline
\textbf{OCT}       & \multicolumn{1}{c|}{0.709 (0.290)} & \multicolumn{1}{c|}{0.747 (0.265)} & \multicolumn{1}{c|}{0.704 (0.260)} & \multicolumn{1}{c|}{0.718 (0.251)} \\ \hline
\end{tabular}}
\vspace{-5mm}
\end{table}

\section{DISCUSSION \& CONCLUSIONS}

In this paper, we present RetStroke, a clinically informed framework that is able to leverage retinal images for stroke detection and prediction. The main advantage of RetStroke is that it is multimodal by nature,  since it incorporates patient's comorbidities, vital-sign measurements, and demographics during training and inference to enhance its accuracy. RetStroke demonstrated a superior performance gain as compared to unimodal models trained on imaging data only, such as OCT and infrared. Furthermore, Retstroke achieve a better performance than an existing foundation model, RetFound \cite{zhou2023foundation}, which consists of around $300$ million parameters and was pre-trained using a significantly larger OCT dataset in the unimodal setting. We investigated incorporating the clinical information in the existing foundation model RetFound and showed that such an adaptation leads to significant improvement in performance. These results highlight the value of the clinical information utilized in our proposed framework. Additionally, our evaluation of RetStroke across diverse patients subgroups including age groups, stroke subtypes and comorbidity indicates its robustness. Overall, RetStroke can be regarded as a novel and unique addition to the field of oculomics to enhance our understanding of an important disease such as stroke.

The unique design of RetStroke provides several advantages that could benefit both clinicians and AI researchers while paving the way for promising research directions. First, RetStroke addresses an unmet clinical need by developing methods for stroke screening using data modalities that are lower in cost, fast to acquire, and independent of stroke symptom onset, thereby improving stroke prevention. Moreover, RetStroke operates on two fronts: stroke risk prediction and detecting lasting effects in the retina. This not only aids in preventing stroke by predicting its risk but also offers insights into retinal changes caused by stroke, which directly impact eye health.

From a technical perspective, RetStroke utilizes a simple and lightweight network, ResNet-18, which is less resource-intensive compared to RetFound. Additionally, its design relies on straightforward fusion techniques, reducing model complexity. Finally, RetStroke leverages simple clinical features that are easy to collect and have been shown to enhance performance. These features could be applied to other relevant use cases.

On the other hand, this study has some limitations that point to promising research directions. First, although our method is compared against external baselines, RetStroke has only been evaluated on an internal private dataset, raising concerns about its generalizability. However, accessing such data is challenging due to healthcare data privacy and regulatory constraints, necessitating further efforts to test the model on external datasets as well as additional baselines. Second, the dataset used for RetStroke development is relatively small (183 patients), which limits the model’s ability to learn stroke-specific features, thereby affecting performance. Addressing this issue requires further research. Possible research directions on this point include training with loss functions that handle class imbalance such as focal loss \cite{lin2017focal}, or applying rare-class oversampling strategies. Third, we define our risk prediction time-frame as a maximum of one year, but longer periods could reveal different insights. Lastly, although RetStroke was developed specifically for stroke, applying it to a broader range of diseases, such as cardiovascular and neurodegenerative conditions, could enhance its clinical utility. Investigating RetStroke in these areas is a pressing need. Overcoming these limitations will contribute to advancing this important field of research.

\addtolength{\textheight}{-12cm}   





\section*{ACKNOWLEDGMENT}
This work was supported by ASPIRE, the technology program management pillar of Abu Dhabi’s Advanced Technology Research Council (ATRC), via the ASPIRE Precision Medicine Research Institute Abu Dhabi (ASPIREPMRIAD) award grant number VRI-20-10, and the NYUAD Center for Artificial Intelligence and Robotics, funded by Tamkeen under the NYUAD Research Institute Award CG010. The research was carried out on the High Performance Computing resources at New York University Abu Dhabi.


\bibliographystyle{ieeetr}
\bibliography{references}

\end{document}